\documentclass[letter]{aa}  

\usepackage{graphicx}
\usepackage{amsmath}  
\usepackage{mathrsfs} 
\usepackage{txfonts}
\usepackage{xcolor}
\usepackage[breaklinks]{hyperref}
\hypersetup{
  colorlinks=true,
    linkcolor=blue,
    citecolor=blue,
    urlcolor=blue}
\usepackage{color}
\usepackage{natbib,twoopt}
\usepackage[hyphenbreaks]{breakurl}
\bibpunct{(}{)}{;}{a}{}{,}             
\definecolor{cobalt}{rgb}{0.06, 0.2, 0.65}
\hypersetup{
  citecolor=cobalt,
  linkcolor=[rgb]{0.8, 0.2, 1.0},
  urlcolor=cobalt,
}
\makeatletter
  \newcommandtwoopt{\citeads}[3][][]{\href{http://adsabs.harvard.edu/abs/#3}%
    {\def\hyper@linkstart##1##2{}%
     \let\hyper@linkend\@empty\citealp[#1][#2]{#3}}}
  \newcommandtwoopt{\citepads}[3][][]{\href{http://adsabs.harvard.edu/abs/#3}%
    {\def\hyper@linkstart##1##2{}%
     \let\hyper@linkend\@empty\citep[#1][#2]{#3}}}
  \newcommandtwoopt{\citetads}[3][][]{\href{http://adsabs.harvard.edu/abs/#3}%
    {\def\hyper@linkstart##1##2{}%
     \let\hyper@linkend\@empty\citet[#1][#2]{#3}}}
  \newcommandtwoopt{\citeyearads}[3][][]%
    {\href{http://adsabs.harvard.edu/abs/#3}
    {\def\hyper@linkstart##1##2{}%
     \let\hyper@linkend\@empty\citeyear[#1][#2]{#3}}}
\makeatother

\begin{document}

   \title{Seismic differences between solar magnetic cycles 23 and 24 for low-degree modes}


   \author{R.~A. García\inst{1}
          \and
          S.~N. Breton\inst{2,3}
          \and
          D. Salabert\inst{4}
          \and
          S. C. Tripathy\inst{5}
          \and
          K. Jain\inst{5}
          \and
          S. Mathur\inst{6,7}
          \and
          E. Panetier\inst{3}
          }

   \institute{Universit\'e Paris-Saclay, Universit\'e Paris Cit\'e, CEA, CNRS, AIM, 91191, Gif-sur-Yvette, France\\  
   \email{rgarcia@cea.fr}
    \and
    INAF – Osservatorio Astrofisico di Catania, Via S. Sofia, 78, 95123 Catania, Italy
    \and
    Universit\'e Paris Cit\'e, Universit\'e Paris-Saclay, CEA, CNRS, AIM, 91191, Gif-sur-Yvette, France  
    \and
    Universit\'e C\^ote d'Azur, Observatoire de la C\^ote d'Azur, CNRS, Laboratoire Lagrange, France    
    \and
    National Solar Observatory, 3665 Discovery Drive, Boulder, CO 80303, USA
    \and
    Instituto de Astrof\'isica de Canarias (IAC), E-38205 La Laguna, Tenerife, Spain
    Instituto de Astrof\'isica de Canarias (IAC), E-38205 La Laguna, Tenerife, Spain
    \and 
    Universidad de La Laguna (ULL), Departamento de Astrof\'isica, E-38206 La Laguna, Tenerife, Spain
    }

   \date{Received ; accepted}

 
  \abstract
 {Solar magnetic activity follows regular cycles of about 11 years with an inversion of polarity in the poles every $\sim$ 22 years. This changing surface magnetism impacts the properties of the acoustic modes. The acoustic mode frequency shifts are a good proxy of the magnetic cycle. 
   In this Letter we investigate solar magnetic activity cycles 23 and 24 through the evolution of the frequency shifts of low-degree modes ($\ell$ = 0, 1, and 2) in three frequency bands. These bands probe properties between 74 and 1575 km beneath the surface.
   The analysis was carried out using observations from the space instrument Global Oscillations at Low Frequency and the ground-based Birmingham Solar Oscillations Network and Global Oscillation Network Group.
   The frequency shifts of radial modes suggest that changes in the magnetic field amplitude and configuration likely occur near the Sun's surface rather than near its core. The maximum shifts of solar cycle 24 occurred earlier at mid and high latitudes (relative to the equator) and about 1550 km beneath the photosphere. At this depth but near the equator, this maximum aligns with the surface activity but has a stronger magnitude.  At around 74 km deep, the behaviour near the equator mirrors the behaviour at the surface, while at higher latitudes, it matches the strength of cycle 23.}
   


   \maketitle
%


\section{Introduction}


\noindent Main-sequence solar-like stars have a convective envelope where the interaction between convection, rotation, and magnetic fields results in a magnetic activity cycle \citep[e.g.][]{2017LRSP...14....4B}. 
However, the reasons why some stars with similar fundamental properties show very different magnetic activity levels 
are still unclear \citep[e.g.][]{1995ApJ...438..269B,2017PhDT.........3E,2021ApJ...912..127S,2024FrASS..1156379S}. 
Large spectroscopic surveys
, such as the one led at the Mount Wilson Observatory \citep{1978ApJ...226..379W} or at Lowell Observatory \citep{2007AJ....133..862H}, and surveys that use spectropolarimetry \citep[e.g.][]{2014MNRAS.444.3517M} or space-based photometric observations \citep{2014A&A...572A..34G,2014ApJS..211...24M,2014A&A...562A.124M,2017A&A...603A..52R,2019ApJS..244...21S,2021ApJS..255...17S} show a large diversity of magnetic variability \citep[cyclic, flat, or variable stars;][]{1985ARA&A..23..379B}. 
Recent studies even suggest that the magnetic solar cycle might soon stop or change its nature \citep[e.g.][]{2018PhT....71f..70M,2019SoPh..294...88M,2022A&A...658A.144N,2024A&A...684A.156N}.
Thus, it is necessary to keep monitoring the temporal evolution of the Sun with a broad set of observables at different heights, from the corona to the sub-surface layers, to impose tighter constraints on current physical models. 


Over the last 50 years, helioseismology, and more recently asteroseismology, has been able to probe the properties of solar and stellar interiors with high precision \citep[e.g.][]{2019LRSP...16....4G,2021LRSP...18....2C}. 
Surface magnetism perturbs the properties of the acoustic oscillation modes (p modes) in several ways. In particular, p-mode properties are affected by magnetic activity; this was first detected in the Sun \citep[e.g.][]{1989A&A...224..253P,1990Natur.345..322E,1992A&A...255..363A,2003ApJ...595..446J,2010ApJ...711L..84T,2017MNRAS.470.1935H,2022MNRAS.514.3821H} and later in solar-like stars \citep[e.g.][]{2010Sci...329.1032G,2011A&A...530A.127S,2013A&A...550A..32M,2019FrASS...6...46M,2016A&A...589A.118S,2017A&A...598A..77K,2018ApJ...852...46K,2018ApJS..237...17S}. 
Mode frequencies are shifted and correlated with the magnetic cycle, while the mode amplitudes are anti-correlated. These frequency shifts can modify the age estimation of a star like the Sun, with variations of up to 6.5$\%$ between solar minima and maxima \citep{2024A&A...688L..17B}.
Studies focused on the low-activity phase between cycles also suggest 
complicated spatial distributions of magnetic fields beneath the solar surface \citep{2009A&A...504L...1S,2022ApJ...924L..20J}. 

Recently, a detailed seismic analysis that included medium- and high-degree modes collected by the Solar Oscillation Imager/Michelson Doppler Imager \citep[SOI/MDI;][]{1995SoPh..162..129S}, 
the Helioseismic and Magnetic Imager \citep[HMI;][]{2012SoPh..275..207S}, 
and the Global Oscillation Network Group \citep[GONG;][]{HarHil1996} by  \citet{2021ApJ...917...45B} has revealed significant changes in the solar convection zone (and even below), 
indicating possible small changes in the position of the tachocline. Moreover, \citet{2021ApJ...920...49I} studied solar quasi-biennial oscillations \citep[QBOs;][]{1979Natur.278..146S} in the rotation rate residuals
and conclude that QBO-like signals were present at different latitudinal bands, with their amplitudes increasing with depth. 
In addition, \citet{10.1093/mnras/stu1329} showed that the quasi-biennial variability in the solar interior depended on the cycle \citep[see also][]{2022MNRAS.515.2415M}. In addition, a recent study by \citet{2023ApJ...959...16J} found different QBO periods in cycles 23 and 24.




Unfortunately, in the case of the asteroseismology of solar-like stars, only low-degree modes can be detected, and the capabilities to diagnose the impact of the surface magnetic activity at different layers are limited. Nevertheless, using these low-degree p modes, it is still possible to evaluate structural perturbations in the $\sim$3000 km beneath the surface \citep{2012ApJ...758...43B,2015A&A...578A.137S,2018IAUS..340...27J}. In this work we investigated the solar seismic signatures of cycles 23 and 24 (from 1996 to 2020) through the evolution of averaged frequency shifts, $\langle\delta\nu\rangle$, of low-degree modes ($\ell$=0, 1, and 2) in different frequency ranges. This allowed us to study, on one hand, the dependence on latitude because radial modes are more sensitive to mid and high latitudes, while dipolar and quadrupolar sectoral modes ($\ell=|m|$) are more sensitive towards the equator and with increasing $\ell$ (see \citealt{,2004ApJ...610L..65J}, and Fig. 3 in \citealt{2014SSRv..tmp...49C}). On the other hand, by averaging the $\langle\delta\nu\rangle$ in different frequency bands, we were able to study the sensitivity of the magnetic perturbation to depth. 
 In this context, we note that \cite{2016ApJ...828...41S} analysed intermediate- and high-degree modes of GONG data for a part of cycle 23 using different frequency ranges and harmonic degrees and conclude that the frequency shifts behave differently at low and high latitudes.
 
The layout of this Letter is as follows. In Sect. 2 we present the observations and the procedure for extracting the p-mode frequency shifts. In Sect. 3 we present the results and discuss them in Sect. 4.




\section{Observations and data analysis}



We used three different datasets of 
velocity measurements. The first one was collected by the Global Oscillations at Low Frequency (GOLF) instrument\footnote{\url{http://irfu.cea.fr/Phocea/Vie_des_labos/Ast/ast_technique.php?id_ast=3842}} \citep[][]{GabGre1995,GabCha1997} on board the Solar and Heliospheric Observatory \citep[SoHO;][]{1995SoPh..162....1D}. We obtained 24.5 years of continuous data -- from April 11, 1996, until September 4, 2020 -- of Doppler velocity time series following \citet{2005A&A...442..385G} and ensuring a proper timing of the measurements \citep[for more details, see][]{2018A&A...617A.108A,2022A&A...658A..27B}. The second dataset was collected with the Birmingham Solar Oscillation Network \citep[BiSON\footnote{\url{http://bison.ph.bham.ac.uk/portal/timeseries}};][]{2016SoPh..291....1H} 
starting at the same time as GOLF but extending only to April 4, 2020 and corrected following the procedure described in \citet{2014MNRAS.441.3009D}. The third dataset is composed of Doppler velocity measurements 
from the GONG network\footnote{\url{https://nispdata.nso.edu/ftp/TSERIES/vmt/}} \citep{,2021PASP..133j5001J}. The series in this third dataset start on May 7, 1995, and are available through October 9, 2020, with a median duty cycle value of 87$\%$ (varying between 78 and 92$\%$).

Both the GOLF and BiSON datasets are divided into consecutive sub-series of 365 days shifted by 91.25 days, while the GONG dataset comprises 360 days shifted by 72 days. 
Series with a duty cycle below 80, 70, and 56$\%$ respectively for GOLF, GONG, and BiSON were removed.
15 consecutive radial orders between 1800~$\mu$Hz and 3790~$\mu$Hz were fitted (as explained in Appendix~\ref{Ap:extraction}) following a traditional, deterministic, maximum likelihood (ML) algorithm or through the \texttt{apollinaire} module\footnote{Documentation is available at \url{https://apollinaire.readthedocs.io/en/latest/}.} \citep{2022A&A...663A.118B}, which implements a Markov chain Monte Carlo (MCMC) Bayesian sampling of the model parameter posterior distribution \citep[][]{2013PASP..125..306F}. The lowest fitted radial order is $n=12$ for $\ell=0$ and 1, and $n=11$ for $\ell=2$. We did not use $\ell = 3$ modes in our analysis because of their low signal-to-noise ratio in Sun-as-a-star data.  

Temporal variations in mode frequencies, $\langle\delta\nu\rangle$, were defined as the differences between the frequencies observed at a given time and the average of the frequencies of years 1996--1997 (independent series 1 and 5), corresponding to the minimum of activity of cycles 22--23. The frequency-shift errors were computed as\begin{equation}
\sigma(\delta\nu) = \sqrt {\frac{1}{\sum{1/\sigma(\nu)^2}}}
.\end{equation}

\begin{figure}[!t]
\begin{center}
    \includegraphics[width=0.40\textwidth]{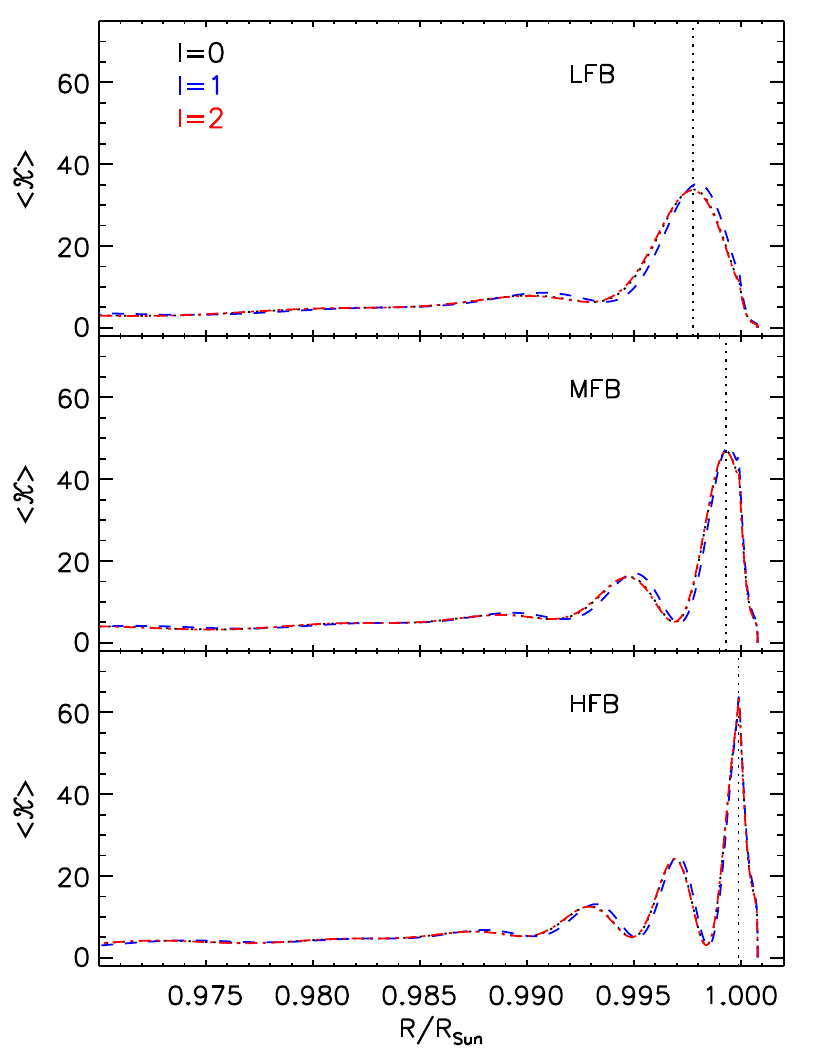}
\end{center}    
    \caption{Integrated sensitivity, $<\mathscr{K}>$, as a function of the relative radius for the LFB, MFB, and HFB (top, middle, and bottom panels, respectively) from the BS05 model in \citet{2005ApJ...626..530B}. Vertical dotted lines indicate the depth of the sensitivity kernels for $\ell$=0 modes.}
    \label{fig:fig_kernel_new}
\end{figure}

Once the frequency shifts were obtained, they were averaged in three different frequency bands following \citet{2012ApJ...758...43B}: (i) the low-frequency band (LFB) from 1800 to 2450 $\mu$Hz; (ii) the medium-frequency band (MFB) from 2450 to 3110 $\mu$Hz; and (iii) the high-frequency band (HFB) from 3110 to 3790 $\mu$Hz. The formal uncertainties resulting from the peak-fitting analysis were used as weights in the computation of the averages of the different degree modes and frequency bands. The resultant $\langle\delta\nu_{\ell=0,1,2}\rangle$ for the three instruments are given in Appendix~\ref{Appendix:Table} (Tables~\ref{Tab:freqshiftsGOLF},~\ref{Tab:freqshiftsBiSON}, and~\ref{Tab:freqshiftsGONG}).
Using model  BS05 from \citet{2005ApJ...626..530B}, we computed the average sensitivity for each degree in each frequency band.  The results are shown in Table~\ref{Tab:Kernels} and Fig.~\ref{fig:fig_kernel_new}. Since the kernels for individual ($n,\ell$) modes have different depth sensitivities, we calculated these values by averaging them over all the $n$ and $\ell$ values in each selection taken from the observations. Our kernels are slightly different from those in \citet{2012ApJ...758...43B} and \citet{2015A&A...578A.137S} because the modes averaged in each frequency band are not the same. For example, \citet{2012ApJ...758...43B} used the same number of overtones (i.e. four) in all three frequency ranges, while we used five consecutive radial orders in each category. This led to the use of a different set of modes in this study and hence a different depth sensitivity.


Finally, because we are interested in comparing solar cycles 23 and 24, we removed the contribution of the QBO \citep[e.g.][]{2010ApJ...718L..19F, 2011ApJ...739....6J, 2012A&A...539A.135S} by smoothing the data with a boxcar of 2.5 years. To reduce any border effects associated with the filtering, we extended the $\langle\delta\nu\rangle$ by half of the filter width, assuming them to be symmetric with respect to the beginning and the end of the series.

In addition, mean values of daily measurements of the 10.7 cm radio flux\footnote{The 10.7 cm radio flux data are available from the LASP Interactive Solar Irradiance Datacenter at \url{https://lasp.colorado.edu/lisird/data/penticton_radio_flux/}.}, $F_{10.7}$, were used as a proxy of the solar surface activity since it has been shown that this magnetic activity proxy is the best correlated with the mean frequency shifts \citep[e.g.][]{2009ApJ...695.1567J,2017MNRAS.470.1935H}. The same 2.5-a-boxcar filter was applied to remove the QBO contribution from this proxy. Using the filtered $F_{10.7}$, we determined the reference dates for the maximum of cycles 23 and 24 and the minimum in between, respectively March 2001, January 2014, and July 2008.

\begin{table}[htp]
\caption{Depths from the surface of the sensitivity kernels.}
\begin{center}
\begin{tabular}{ccccc}
\hline
\hline

$\nu$  & Depth & Depth & Depth & Basu et al. \\
band    & ($\ell$=0)  & ($\ell$=1) &  ($\ell$=2) & (2012) \\
\hline
LFB &  0.99777 R$_\odot$   &   0.99798 R$_\odot$      &     0.99774 R$_\odot$  &  0.9963 R$_\odot$    \\
&  (1550 km)   &   (1402 km)     & (1575 km)    &  (2576 km)   \\
MFB & 0.99932 R$_\odot$   &  0.99941 R$_\odot$     &     0.99930 R$_\odot$     & 0.9981 R$_\odot$  \\
& (474 km)& (412 km) & (487 km)    & (1323 km)\\
HFB &0.99989 R$_\odot$       & 0.99989 R$_\odot$      &    0.99989 R$_\odot$    & 0.9989 R$_\odot$  \\
& (74 Km) & (74 km) & (75 km)   & (766 km) \\
\hline
\end{tabular}
\end{center}
\label{Tab:Kernels}
\flushleft {\bf Notes.} The depths were computed with the BS05 model from \citet{2005ApJ...626..530B} for modes $\ell$ = 0, 1, and 2 and the three frequency bands. The last column indicates the depth computed by \citet{2012ApJ...758...43B}.
\end{table}%

By comparing the mode parameters from MDI and HMI, which used two different spectral lines (Ni I 676.8 nm and Fe I 617.3 nm, respectively), \cite{2018SoPh..293...29L} found that differences in frequencies and splitting coefficients were not significant. In addition, no differences were found when comparing the temporal evolution of p-mode parameters of GOLF and BiSON \citep{2004ApJ...604..969J} except for the mode asymmetry \citep{JimCha2007}. Thus, we conclude that our results should not be affected by the fact that they are based on the study of the temporal evolution of  $\langle\delta\nu\rangle$ involving different instruments observing the signals at slightly different heights.



\begin{figure*}[!htb]
\begin{center}
    \includegraphics[width=0.82\textwidth,trim=90 240 130 100, clip]{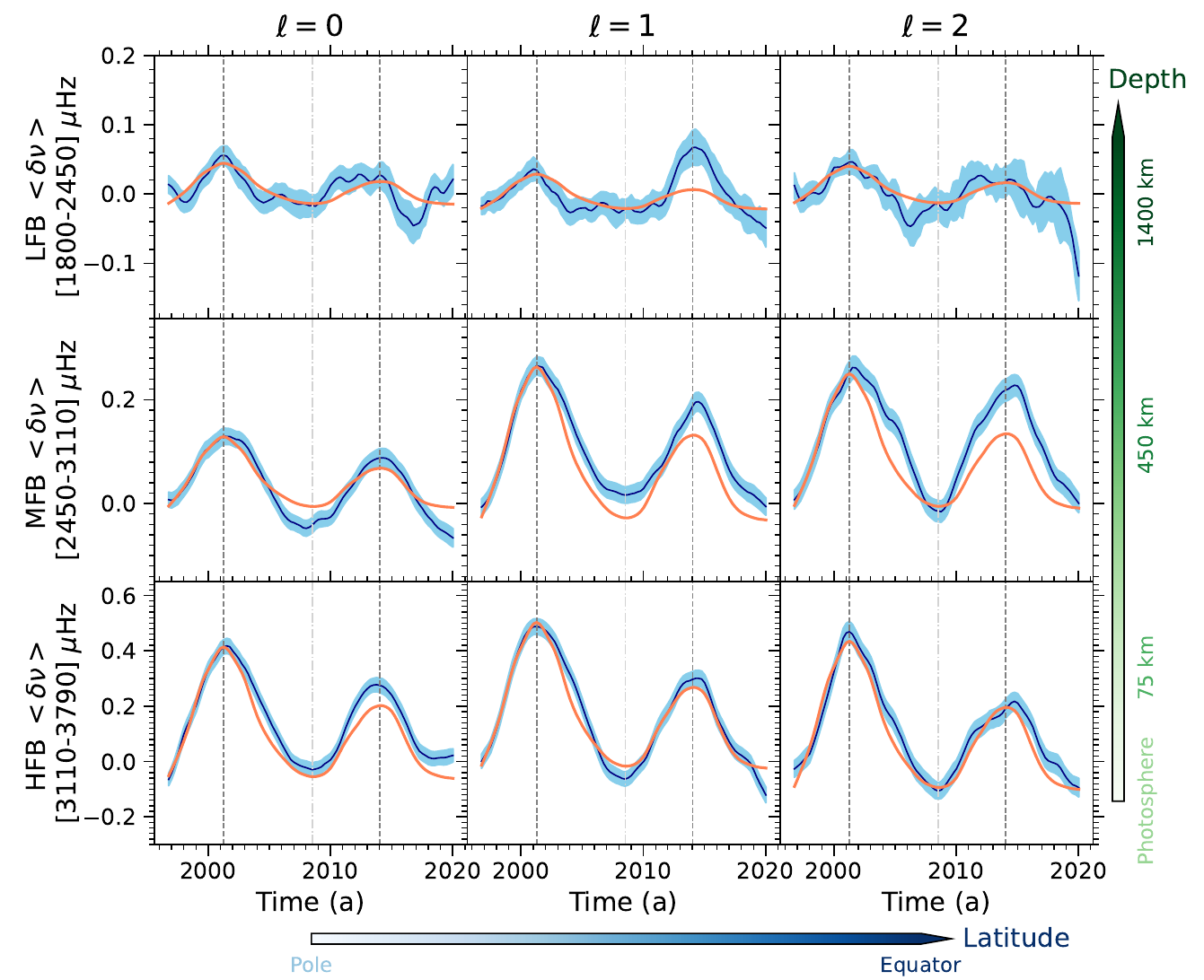} 
\end{center}    
    \caption{Temporal evolution of the averaged and smoothed GOLF frequency shifts ($\langle\delta\nu\rangle$, dark blue lines) of $\ell$=0, 1, and 2 modes (left, central, and right panels respectively). The top, middle, and bottom rows correspond to the three frequency bands considered in the analysis: 1800-2450 (LFB), 2450-3110 (MFB), and 3110-3790 $\mu$Hz (HFB), respectively.  Light blue regions represent the errors in $\langle\delta\nu\rangle$. Orange lines depict the scaled smoothed $F_{10.7}$ magnetic proxy. Vertical dashed and dot-dashed lines depict the maxima of cycles 23 and 24 and the minimum in between. The horizontal blue arrow indicates the change in sensitivity from the pole to the equator, and the vertical green arrow depicts the maximum sensitivity to the depth of the subsurface layers.}
    \label{fig:fig1}
\end{figure*}

\section{Results: Temporal evolution of GOLF frequency shifts}
The temporal evolution of the 2.5-a-boxcar smoothed $\langle\delta\nu\rangle$ from GOLF is shown in Fig.~\ref{fig:fig1}. Each column corresponds to a different mode ($\ell$=0, 1, and 2), and each row corresponds to the aforementioned frequency bands. A figure containing the non-smoothed $\langle\delta\nu\rangle$ is shown in Appendix~\ref{Appendix:App_QBO}. In each panel, the  $F_{10.7}$ radio flux is scaled to the $\langle\delta\nu\rangle$ using the rising phase of cycle 23. For the sake of clarity and because the qualitative results are the same for the three instruments, only the GOLF results are presented. In Appendix \ref{Appendix:BiSON_GONG}, results obtained with the other two datasets are compared and discussed. 

Overall, the $\langle\delta\nu\rangle$ follow the evolution of the $F_{10.7}$, but there are some notable differences:

\begin{itemize}
\item The rising phase of cycle 24 in the LFB $\langle\delta\nu_{\ell=0}\rangle$ is reached in 2012, two years earlier than $F_{10.7}$. 
\item Cycle 24 is known to have a lower amplitude than cycle 23, as shown for example by \citet{2022MNRAS.514.3821H} and by the  $F_{10.7}$. 
In this respect, $\langle\delta\nu\rangle$ modulations have surprisingly large amplitudes in cycle~24 for some frequency bands and degrees, $\ell$.
The amplitude of the LFB $\langle\delta\nu_{\ell=1}\rangle$ of cycle 24 is nearly twice that of cycle 23, while for the MFB $\langle\delta\nu_{\ell=2}\rangle$,
the amplitude of cycle 24 is comparable to that of cycle 23. In all of these cases, the amplitude of cycle 24 is clearly higher than that of the $F_{10.7}$. 
The MFB $\langle\delta\nu_{\ell=0}\rangle$ reaches its minimum between the two cycles around a year earlier than the $F_{10.7}$.
\item The MFB $\langle\delta\nu_{\ell=1}\rangle$ and HFB $\langle\delta\nu_{\ell=0}\rangle$ are both greater than $F_{10.7}$ in cycle 24.
\item With larger and larger error bars in the LFB $\langle\delta\nu_{\ell=2}\rangle$ since 2005, due to the lower counting rates of the GOLF instrument, we cannot extract information on cycle~24.
\item For the remaining three cases, the MFB $\langle\delta\nu_{\ell=0}\rangle$ and the HFB $\langle\delta\nu_{\ell=1}\rangle$ and $\langle\delta\nu_{\ell=2}\rangle$, they closely follow $F_{10.7}$.
\end{itemize}


\section{Discussion}
Recent seismic analyses of averaged frequency shifts for both low-degree modes \citep{2022MNRAS.514.3821H} and medium- and high-degree modes \citep{2022ApJ...924L..20J} show that the influence of the magnetic perturbations on the acoustic modes are different during consecutive solar cycles. Although it is usually assumed that the main perturbations on the mode frequencies are located in the sub-surface layers, it has also been speculated that a fossil magnetic field in the core of the Sun could play a crucial role in the generation of the solar magnetic cycle and thus could also modify the properties of the acoustic modes \citep[e.g.][]{1983Natur.306..670S,2001SoPh..198...51M}. Such magnetic fields have not been observed in main-sequence solar-like stars yet, but they could modify the structure of stellar oscillations \citep[e.g.][]{GooTho1992,2018MNRAS.477.5338L,2021A&A...650A..53B}. They have been observed in the core of red giants \citep{2022Natur.610...43L,2024A&A...688A.184L,2023A&A...670L..16D}.

As discussed in the previous section, the maximum of the $\langle\delta\nu_{\ell=0}\rangle$ in cycle 24 arrives two years earlier in the LFB. This means that something changed between the two cycles at a depth of $\sim$1550 km below the photosphere (see Table~\ref{Tab:Kernels} and Fig. \ref{fig:fig_kernel_new}). The dominant perturbation in the LFB $\langle\delta\nu_{\ell=0}\rangle$ is weighted towards mid to high latitudes, while the maximum of $\langle\delta\nu_{\ell=1}\rangle$ is synchronised with  $F_{10.7}$ except in regions close to the equator, where $\langle\delta\nu_{\ell=1}\rangle$ are the most sensitive. Therefore, in cycle 24, the magnetic perturbation modifying $\langle\delta\nu_{\ell=0}\rangle$ reaches an extended 4-a maximum earlier ($\sim$2010) at high latitudes and at a depth of $\sim$1550 km beneath the surface. This perturbation reaches shallower layers ($>$1402 km) at the same time as the photosphere, with a maximum of around $\sim$2014.
At these mid and high latitudes, just below the photosphere ($\sim$74km), the HFB of $\langle\delta\nu_{\ell=0}\rangle$ shows stronger variations than the surface activity represented by the $F_{10.7}$ in cycle 24. Based on our analysis of the radial modes, we do not find evidence supporting that the changes comes from an hypothetical magnetic field in the core of the Sun as the MFB $\langle\delta\nu_{\ell=0}\rangle$ seems to follow the evolution of $F_{10.7}$ very closely. The fact that all the radial modes reach the centre rules out a change coming from the core of the Sun. Only the upper turning points are different. Thus, the change in the perturbation should be located at the subsurface layers, which likely have a complicated vertical structure. 

Near the solar equator, we see that the LFB of $\langle\delta\nu_{\ell=1}\rangle$ is stronger in cycle 24 than in cycle 23. The perturbation of the dipolar modes at $\sim$ 1402 km is larger in the later cycle. The differences in the $\langle\delta\nu\rangle$ of the ${\ell=1}$ and 2 modes compared to the $F_{10.7}$ in cycle 24 are also large in the MFB. In this cycle, the maximum of the dipolar  $\langle\delta\nu\rangle$ is slightly smaller than in cycle 23, while it is larger for the quadrupolar modes. That means that the perturbation is larger very close to the equator since the difference in the depth is very small ($\sim$~412 versus $\sim$~487 km). Finally, very close to the surface ($\sim$~74 -- 75 km), the behaviour of {\color{purple} $\langle\delta\nu\rangle$} is almost identical to the surface activity.
 Despite the different frequency bands and mode degrees, we find similar qualitative results regarding the dependence of the progression of solar cycle 23 to those obtained by \cite{2016ApJ...828...41S}.

\section{Conclusions}

Similar to previous studies using intermediate- and high-degree modes \citep[e.g.][and references therein]{Howe_2022}, the present study allowed us to trace how the solar-cycle-induced perturbations evolve with time, latitude, and depth using low-degree modes during solar cycles 23 and 24, and with three different instruments, GOLF, BiSON, and GONG.
Our first conclusion is that the temporal evolution of the radial-mode frequency shifts suggests that these changes most likely did not occur in the solar core but rather close to the surface. The second conclusion is that the maximum shifts of cycle 24 seem to arrive earlier at mid and high latitudes compared to cycle 23 and at a depth of around 1550~km. At this depth and near the equator, the maximum of cycle 24 is synchronised with the surface, but the magnitude is stronger than at the surface. Finally, in a band around 74~km beneath the surface and near the equator, the behaviour is similar to that at the surface, while at higher latitudes it matches the strength of cycle 23. The results obtained with the three instruments are in close agreement in most of the cases. However, as shown in Appendix~\ref{Appendix:Table}, the main difference is the presence of a clear QBO pattern in the GONG LFB of $\langle\delta\nu_{\ell=0}\rangle$ that is not seen in the integrated-light instruments. The smoothed frequency shift also behaves differently, closely following the $F_{10.7}$.












\section{Data availability}
Tables \ref{Tab:freqshiftsGOLF},~\ref{Tab:freqshiftsBiSON}, and~\ref{Tab:freqshiftsGONG} are only available in electronic form at the CDS via anonymous ftp to cdsarc.u-strasbg.fr (130.79.128.5) or via http://cdsweb.u-strasbg.fr/cgi-bin/qcat?J/A+A/.

\begin{acknowledgements}
We thank J. Ballot and F. P\'erez Hern\'andez for useful comments and discussions. K.J. thanks Sarbani Basu for providing the standard solar model values. The GOLF instrument on board SoHO is a cooperative effort of many individuals, to whom we are indebted. SoHO is a project of international collaboration between ESA and NASA. BiSON is funded by the UK Science and Technology Facilities Council (STFC). This work also utilises GONG data obtained by the NSO Integrated Synoptic Program, managed by the National Solar Observatory, which is operated by the Association of Universities for Research in Astronomy (AURA), Inc. under a cooperative agreement with the National Science Foundation and with contribution from the National Oceanic and Atmospheric Administration. The GONG network of instruments is hosted by the Big Bear Solar Observatory, High Altitude Observatory, Learmonth Solar Observatory, Udaipur Solar Observatory, Instituto de Astrof\'{\i}sica de Canarias, and Cerro Tololo Interamerican Observatory.
R.A.G., S.N.B., and E.P. acknowledge the support from the GOLF and PLATO Centre National D'{\'{E}}tudes Spatiales grants. 
S.N.B acknowledges support from PLATO ASI-INAF agreement n.~2015-019-R.1-2018.
S.M.\ acknowledges support by the Spanish Ministry of Science and Innovation with the Ramon y Cajal fellowship number RYC-2015-17697 and the grant number PID2019-107187GB-I00, and through AEI under the Severo Ochoa Centres of Excellence Programme 2020--2023 (CEX2019-000920-S). S.C.T. and K.J. acknowledge partial funding support from the NASA DRIVE Science Center COFFIES Phase II CAN 80NSSC22M0162. 

\\
\textit{Software:} AstroPy \citep{astropy:2013,astropy:2018}, Matplotlib \citep{matplotlib}, NumPy \citep{numpy}, SciPy \citep{scipy}, pandas \citep{mckinney-proc-scipy-2010_pandas,reback2020pandas}, emcee \citep{2013PASP..125..306F}, \texttt{apollinaire} \citep{2022A&A...663A.118B}, 
Interactive Data Language (IDL\footnote{\url{https://www.nv5geospatialsoftware.com/docs/home.html}}).

\end{acknowledgements}

%
%
\bibliographystyle{aa_url.bst}
\bibliography{BIBLIO.bib} 

\begin{appendix} 
\section{Extraction of the p-mode parameters}
\label{Ap:extraction}

We extracted the parameters of the individual p modes using three different methods that we describe below.


\subsection{GOLF: MCMC fitting}


 We used the Lorentzian-profile p-mode model as described in Eq.~\ref{eq:mlemodel} to perform an MCMC analysis of the GOLF power spectra with the \texttt{apollinaire} library \citep{2022A&A...663A.118B} which implements the ensemble sampler provided by the \texttt{emcee} library \citep{2013PASP..125..306F} for asteroseismic analyses. Given a set of mode parameters, $\theta$, the MCMC process provides a sampling of the posterior probability, $p (\theta | \mathbf{S_x})$:
\begin{equation}
    p (\theta | \mathbf{S_x}) = p (\mathbf{S_x} | \theta) p (\theta) \;,
\end{equation}
where $p (\mathbf{S_x} | \theta)$ is the likelihood function and $p(\theta)$ the prior distribution of $\theta$. 
The marginal likelihood $p (\mathbf{S_x})$ is omitted in the formula.
In order to consider prior uniform distributions while preserving non-informative prior for every sampled parameter \citep{Benomar2009a}, we choose to sample the logarithms $\ln \Gamma_{n,\ell}$ and $\ln H_{n,\ell}$ rather than sampling directly $\Gamma_{n,\ell}$ and $H_{n,\ell}$.
The extracted mode parameters are taken as the median of the sampled distribution. The uncertainty we consider is the largest value between the absolute difference of the median value and the 16$\rm ^{th}$ centile value of the distribution and the absolute difference of the median value and the 84$\rm ^{th}$ centile value of the distribution. A detailed comparison of the ML and MCMC fitting results based on GOLF data can be found in  \citet{2022A&A...663A.118B}.

\subsection{BiSON: ML fitting \label{sec:mle_golf}}
The power spectrum of each time series was fitted assuming a $\chi^2$ with two degrees of freedom likelihood to estimate the mode parameters of the $\ell$=0, 1, 2, and 3 modes as described in \citet{2007A&A...463.1181S}. Each mode component of radial order $n$, angular degree $l$, and azimuthal order $m$ was parametrised with an asymmetric Lorentzian profile \citep{1998ApJ...505L..51N}, as
\begin{equation}
{\cal L}_{n,l,m}(\nu) = H_{n,l} \frac{(1+b_{n,l} x_{n,l})^2+b_{n,l}^2}{1+ x_{n,l}^2},
\label{eq:mlemodel}
\end{equation} 
\noindent
where $x_{n,l} = 2(\nu-\nu_{n,l})/\Gamma_{n,l}$, and $\nu_{n,l}$, $\Gamma_{n,l}$, and $H_{n,l}$ represent the mode frequency, linewidth, and height of the spectral density, respectively. The peak asymmetry is described by the parameter $b_{n,l}$.
Because of their close proximity in frequency, modes are fitted in pairs (i.e. $l=2$ with 0, and $l=3$ with 1). While each mode parameter within a pair of modes is free, the peak asymmetry is set to be the same within pairs of modes. The $l = 4$ and 5 modes were included in the fitted model when they are present within the fitting window, which is proportional to the mode linewidth (i.e. frequency dependent). An additive parameter $B$ is added to the fitted profile representing a constant background noise in the fitting window. Since SoHO and BiSON observe the Sun equatorward, only the $l+|m|$ even components are visible in Sun-as-a-star observations. The amplitude ratios between the $l=0,1,2$, and 3 modes and the $m$-height ratios of the $l=2$ and 3 multiplets correspond to those calculated in \citet{2011A&A...528A..25S}.  
Finally, the mode parameters were extracted by maximizing the likelihood function, the power spectrum statistics being described by a $\chi^2$ distribution with two degrees of freedom.
The natural logarithms of the mode height, linewidth, and background noise were varied resulting in normal distributions. The formal uncertainties in each parameter were then derived from the inverse Hessian matrix. \\


\subsection{GONG: ML fitting}
 We used $p$-mode frequencies corresponding to the individual multiplets ($\nu_{n, \ell, m}$) obtained from the  GONG time series, where $\nu$ is the frequency,  $n$ is the radial order and $m$ is the azimuthal order, running from $-\ell$ to $+\ell$.  The multiplets were estimated from the $m-\nu$ power spectra constructed from time series spanning 360 days (10 GONG months, where 1 GONG month corresponds to 36 days)  with a spacing of 72 days between two consecutive data sets.  The time series were processed through the standard GONG peak-fitting algorithm to compute the power spectra based on the multi-taper spectral analysis coupled with a fast Fourier transform \citep{1999ApJ...519..407K}. Finally, Lorentzian profiles were used to fit the peaks in the $m-\nu$ spectra using a minimisation scheme guided by an initial guess table \citep{1990ApJ...364..699A}.
In order to compare the mode frequencies for low-degree modes obtained from GOLF instrument, we averaged the multiplets over all $m$ values for a given $\ell$ and $n$ value to obtain m-averaged mode frequencies (i.e. quadruple modes were fitted for m $\pm$2 including 0 and dipole using m $\pm$ 1).


\section{Computed averaged frequency shifts}
\label{Appendix:Table}
In Tables~\ref{Tab:freqshiftsGOLF},~\ref{Tab:freqshiftsBiSON}, and~\ref{Tab:freqshiftsGONG}, the computed averaged frequency shifts $\langle\delta\nu\rangle$ for modes $\ell=$ 0, 1, and 2 are shown respectively for GOLF, BiSON, and GONG.

\begin{table*}[htp]
\tiny
\caption{GOLF $\langle\delta\nu\rangle$ for modes $\ell$ = 0, 1, and 2 for the three frequency bands, LFB, MFB, and HFB.}
\begin{center}
\begin{tabular}{cccccccccc}
\hline
\hline
    & & & & & GOLF & & & & \\
\hline  
Time $[a]$ & & $\langle\delta\nu\rangle$ LFB $[\mu$Hz$]$& & & $\langle\delta\nu\rangle$ MFB $[\mu$Hz$]$& & &  $\langle\delta\nu\rangle$ HFB $[\mu$Hz$]$&  \\
        & $\ell$=0  & $\ell$=1 &  $\ell$=2 & $\ell$=0  & $\ell$=1 &  $\ell$=2 & $\ell$=0  & $\ell$=1 &  $\ell$=2 \\
\hline
1996.770 & 0.014 $\pm$ 0.021 & -0.017 $\pm$ 0.020 & 0.012 $\pm$ 0.030 & & ... & & -0.065 $\pm$ 0.037 & 0.001 $\pm$ 0.042 & -0.027 $\pm$ 0.050 \\ 
1997.019 & -0.003 $\pm$ 0.020 & -0.002 $\pm$ 0.021 & 0.014 $\pm$ 0.025 & & ... & & 0.004 $\pm$ 0.038 & -0.050 $\pm$ 0.041 & -0.033 $\pm$ 0.049 \\ 
1997.268 & -0.020 $\pm$ 0.022 & -0.017 $\pm$ 0.021 & 0.030 $\pm$ 0.027 & & ... & & 0.020 $\pm$ 0.038 & -0.029 $\pm$ 0.039 & 0.009 $\pm$ 0.051 \\ 
1997.518 & -0.006 $\pm$ 0.024 & -0.001 $\pm$ 0.022 & 0.001 $\pm$ 0.027 & & ... & & 0.062 $\pm$ 0.038 & -0.018 $\pm$ 0.040 & 0.016 $\pm$ 0.052 \\ 
1997.770 & -0.018 $\pm$ 0.024 & 0.030 $\pm$ 0.026 & -0.010 $\pm$ 0.028 & & ... & & 0.073 $\pm$ 0.039 & -0.001 $\pm$ 0.040 & 0.031 $\pm$ 0.054 \\ 
1998.019 & -0.007 $\pm$ 0.025 & -0.002 $\pm$ 0.023 & -0.039 $\pm$ 0.028 & & ... & & 0.054 $\pm$ 0.041 & 0.087 $\pm$ 0.043 & 0.038 $\pm$ 0.052 \\ 
1998.268 & -0.003 $\pm$ 0.023 & -0.020 $\pm$ 0.027 & -0.074 $\pm$ 0.027 & & ... & & 0.081 $\pm$ 0.041 & 0.124 $\pm$ 0.042 & 0.011 $\pm$ 0.049 \\ 
1998.518 & -0.014 $\pm$ 0.026 & 0.004 $\pm$ 0.026 & -0.070 $\pm$ 0.027 & & ... & & 0.125 $\pm$ 0.042 & 0.178 $\pm$ 0.044 & 0.018 $\pm$ 0.050 \\ 
1998.770 & -0.045 $\pm$ 0.027 & -0.042 $\pm$ 0.023 & 0.036 $\pm$ 0.023 & & ... & & 0.198 $\pm$ 0.042 & 0.240 $\pm$ 0.045 & 0.072 $\pm$ 0.048 \\ 
1999.019 & -0.027 $\pm$ 0.027 & -0.027 $\pm$ 0.022 & 0.061 $\pm$ 0.021 & & ... & & 0.270 $\pm$ 0.042 & 0.381 $\pm$ 0.047 & 0.057 $\pm$ 0.054 \\ 
...& & & & & & & & \\
\hline
\hline
\end{tabular}
\end{center}
\label{Tab:freqshiftsGOLF}
\flushleft {\bf Notes.}  Time is given in fractional years and frequency shifts and errors are in $\mu$Hz. The complete table with all significant digits is available online in a machine-readable format.
\end{table*}

\begin{table*}[htp]
\tiny
\caption{BiSON $\langle\delta\nu\rangle$ for modes $\ell$ = 0, 1, and 2 for the three frequency bands, LFB, MFB, and HFB.}
\begin{center}
\begin{tabular}{cccccccccc}
\hline
\hline
    & & & & & BiSON & & & & \\
\hline  
Time $[a]$ & & $\langle\delta\nu\rangle$ LFB $[\mu$Hz$]$& & & $\langle\delta\nu\rangle$ MFB $[\mu$Hz$]$& & &  $\langle\delta\nu\rangle$ HFB $[\mu$Hz$]$&  \\
        & $\ell$=0  & $\ell$=1 &  $\ell$=2 & $\ell$=0  & $\ell$=1 &  $\ell$=2 & $\ell$=0  & $\ell$=1 &  $\ell$=2 \\
\hline
1996.770 & -0.002 $\pm$ 0.035 & -0.026 $\pm$ 0.034 & -0.001 $\pm$ 0.027 & & ... & & -0.072 $\pm$ 0.048 & 0.027 $\pm$ 0.051 & -0.060 $\pm$ 0.052 \\ 
1997.019 & 0.008 $\pm$ 0.033 & -0.025 $\pm$ 0.035 & -0.005 $\pm$ 0.025 & & ... & & -0.017 $\pm$ 0.050 & -0.056 $\pm$ 0.051 & -0.066 $\pm$ 0.050 \\ 
1997.268 & -0.008 $\pm$ 0.034 & -0.026 $\pm$ 0.033 & 0.013 $\pm$ 0.029 & & ... & & -0.003 $\pm$ 0.052 & -0.046 $\pm$ 0.051 & -0.023 $\pm$ 0.054 \\ 
1997.518 & -0.016 $\pm$ 0.036 & -0.001 $\pm$ 0.034 & -0.003 $\pm$ 0.026 & & ... & & 0.073 $\pm$ 0.052 & -0.023 $\pm$ 0.050 & 0.039 $\pm$ 0.056 \\ 
1997.770 & 0.002 $\pm$ 0.038 & 0.031 $\pm$ 0.037 & 0.001 $\pm$ 0.028 & & ... & & 0.087 $\pm$ 0.053 & -0.028 $\pm$ 0.052 & 0.070 $\pm$ 0.056 \\ 
1998.019 & -0.014 $\pm$ 0.041 & 0.010 $\pm$ 0.036 & -0.022 $\pm$ 0.027 & & ... & & 0.112 $\pm$ 0.056 & 0.111 $\pm$ 0.051 & 0.066 $\pm$ 0.056 \\ 
1998.268 & 0.025 $\pm$ 0.041 & 0.008 $\pm$ 0.038 & -0.024 $\pm$ 0.028 & & ... & & 0.189 $\pm$ 0.055 & 0.189 $\pm$ 0.050 & 0.118 $\pm$ 0.054 \\ 
1998.518 & 0.039 $\pm$ 0.041 & 0.034 $\pm$ 0.037 & -0.047 $\pm$ 0.031 & & ... & & 0.255 $\pm$ 0.055 & 0.251 $\pm$ 0.051 & 0.135 $\pm$ 0.053 \\ 
1998.770 & -0.014 $\pm$ 0.041 & 0.036 $\pm$ 0.036 & -0.015 $\pm$ 0.030 & & ... & & 0.294 $\pm$ 0.053 & 0.309 $\pm$ 0.051 & 0.199 $\pm$ 0.052 \\ 
1999.019 & -0.022 $\pm$ 0.040 & 0.044 $\pm$ 0.035 & 0.034 $\pm$ 0.028 & & ... & & 0.301 $\pm$ 0.051 & 0.324 $\pm$ 0.051 & 0.232 $\pm$ 0.054 \\ 
...& & & & & & & & \\
\hline
\hline
\end{tabular}
\end{center}
\label{Tab:freqshiftsBiSON}
\flushleft {\bf Notes.} Time is given in fractional years and frequency shifts and errors are in $\mu$Hz. The complete table with all significant digits is available online in a machine-readable format.
\end{table*}

\begin{table*}[htp]
\tiny
\caption{GONG $\langle\delta\nu\rangle$ for modes $\ell$ = 0, 1, and 2 for the three frequency bands, LFB, MFB, and HFB.}
\begin{center}
\begin{tabular}{cccccccccc}
\hline
\hline
    & & & & & GONG & & & & \\
\hline  
Time $[a]$ & & $\langle\delta\nu\rangle$ LFB $[\mu$Hz$]$& & & $\langle\delta\nu\rangle$ MFB $[\mu$Hz$]$& & &  $\langle\delta\nu\rangle$ HFB $[\mu$Hz$]$&  \\
        & $\ell$=0  & $\ell$=1 &  $\ell$=2 & $\ell$=0  & $\ell$=1 &  $\ell$=2 & $\ell$=0  & $\ell$=1 &  $\ell$=2 \\
\hline
1997.022 & -0.007 $\pm$ 0.028 & 0.011 $\pm$ 0.068 & 0.030 $\pm$ 0.053 & & ... & & 0.026 $\pm$ 0.040 & -0.042 $\pm$ 0.066 & -0.021 $\pm$ 0.075 \\ 
1997.219 & -0.030 $\pm$ 0.028 & 0.006 $\pm$ 0.066 & 0.026 $\pm$ 0.055 & & ... & & 0.037 $\pm$ 0.041 & -0.075 $\pm$ 0.070 & 0.004 $\pm$ 0.075 \\ 
1997.416 & -0.012 $\pm$ 0.028 & 0.060 $\pm$ 0.061 & 0.039 $\pm$ 0.051 & & ... & & 0.064 $\pm$ 0.042 & -0.070 $\pm$ 0.071 & 0.021 $\pm$ 0.075 \\ 
1997.614 & -0.008 $\pm$ 0.029 & 0.058 $\pm$ 0.062 & 0.046 $\pm$ 0.050 & & ... & & 0.082 $\pm$ 0.042 & -0.054 $\pm$ 0.074 & 0.039 $\pm$ 0.078 \\ 
1997.811 & 0.013 $\pm$ 0.029 & 0.006 $\pm$ 0.063 & 0.038 $\pm$ 0.052 & & ... & & 0.057 $\pm$ 0.043 & -0.077 $\pm$ 0.075 & 0.088 $\pm$ 0.077 \\ 
1998.008 & 0.039 $\pm$ 0.027 & -0.022 $\pm$ 0.068 & 0.013 $\pm$ 0.054 & & ... & & 0.044 $\pm$ 0.045 & -0.019 $\pm$ 0.076 & 0.103 $\pm$ 0.078 \\ 
1998.205 & 0.048 $\pm$ 0.030 & -0.186 $\pm$ 0.055 & -0.009 $\pm$ 0.054 & & ... & & 0.104 $\pm$ 0.046 & -0.004 $\pm$ 0.079 & 0.165 $\pm$ 0.079 \\ 
1998.403 & 0.054 $\pm$ 0.028 & -0.006 $\pm$ 0.069 & 0.004 $\pm$ 0.058 & & ... & & 0.149 $\pm$ 0.045 & 0.076 $\pm$ 0.081 & 0.218 $\pm$ 0.079 \\ 
1998.600 & 0.030 $\pm$ 0.031 & 0.042 $\pm$ 0.062 & -0.002 $\pm$ 0.058 & & ... & & 0.161 $\pm$ 0.046 & 0.101 $\pm$ 0.078 & 0.251 $\pm$ 0.078 \\ 
1998.797 & 0.018 $\pm$ 0.028 & 0.066 $\pm$ 0.056 & -0.002 $\pm$ 0.053 & & ... & & 0.203 $\pm$ 0.046 & 0.179 $\pm$ 0.078 & 0.243 $\pm$ 0.077 \\ 
...& & & & & & & & \\
\hline
\hline
\end{tabular}
\end{center}
\label{Tab:freqshiftsGONG}
\flushleft {\bf Notes.} Time is given in fractional years and frequency shifts and errors are in $\mu$Hz. The complete table with all significant digits is available online in a machine-readable format.
\end{table*}

\section{Non-smoothed GOLF results showing the QBO}
\label{Appendix:App_QBO}

\begin{figure*}[!hb]
\begin{center}
    \includegraphics[width=1\textwidth,trim=80 220 130 100, clip]{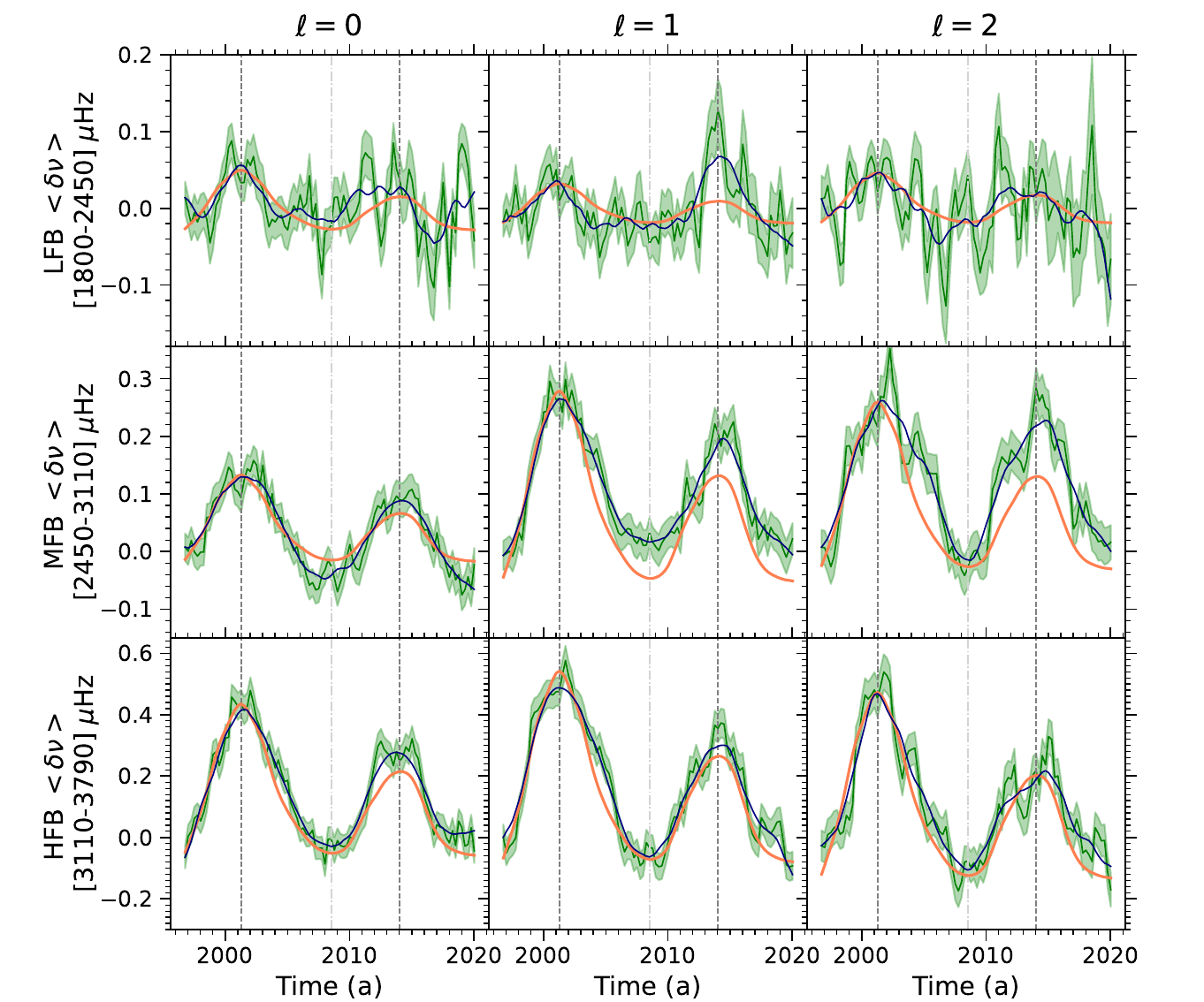}
\end{center}    
    \caption{Temporal evolution of the averaged but not smoothed GOLF frequency shifts (dark green lines). Errors are shown as light green regions. As a reference, the dark blue line represents the smoothed frequency shifts as shown in Fig.~\ref{fig:fig1}. Orange and vertical grey lines have the same meaning as in Fig.~\ref{fig:fig1}.}
    \label{fig:fig_with_QBO}
\end{figure*}

Figure~\ref{fig:fig_with_QBO} shows the same smoothed frequency shifts and $F_{10.7}$ shown in Fig.~\ref{fig:fig1} but including the non-filtered frequency shifts in dark blue. The behaviour of these frequency shifts is dominated by the presence of the QBO.

\section{Results for BiSON and GONG}
\label{Appendix:App_BISON_GONG}
\begin{figure*}[!htb]
\begin{center}
    \includegraphics[width=1.\textwidth,trim=80 220 130 100, clip]{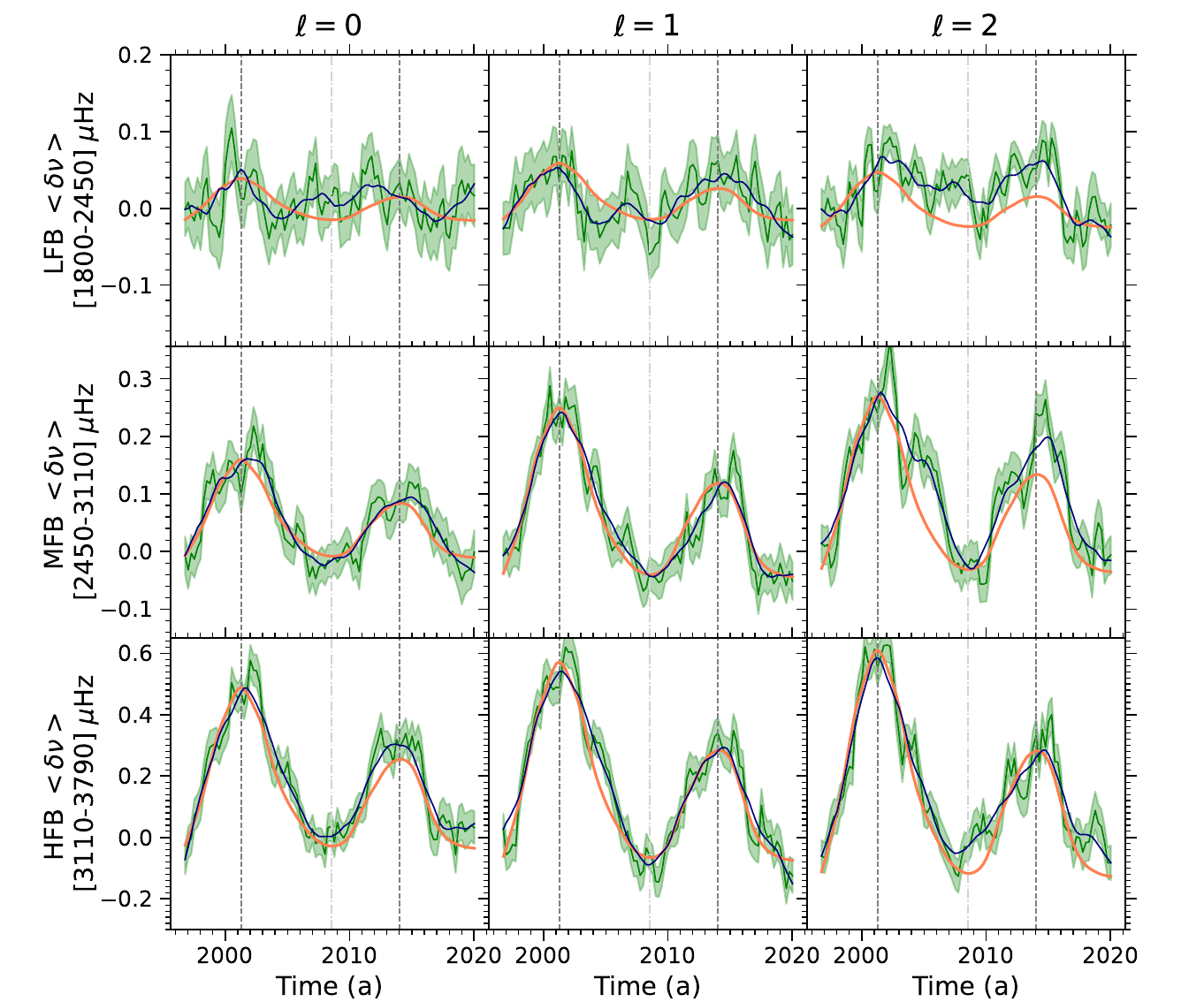}
\end{center}    
    \caption{Temporal evolution of averaged but not smoothed BiSON frequency shifts (dark green lines) with errors represented by the light green regions. Same legend as in Fig.~\ref{fig:fig_with_QBO}.}
    \label{fig:fig_BiSON}
\end{figure*}

\begin{figure*}[!htb]
\begin{center}
    \includegraphics[width=1.\textwidth,trim=80 220 130 100, clip]{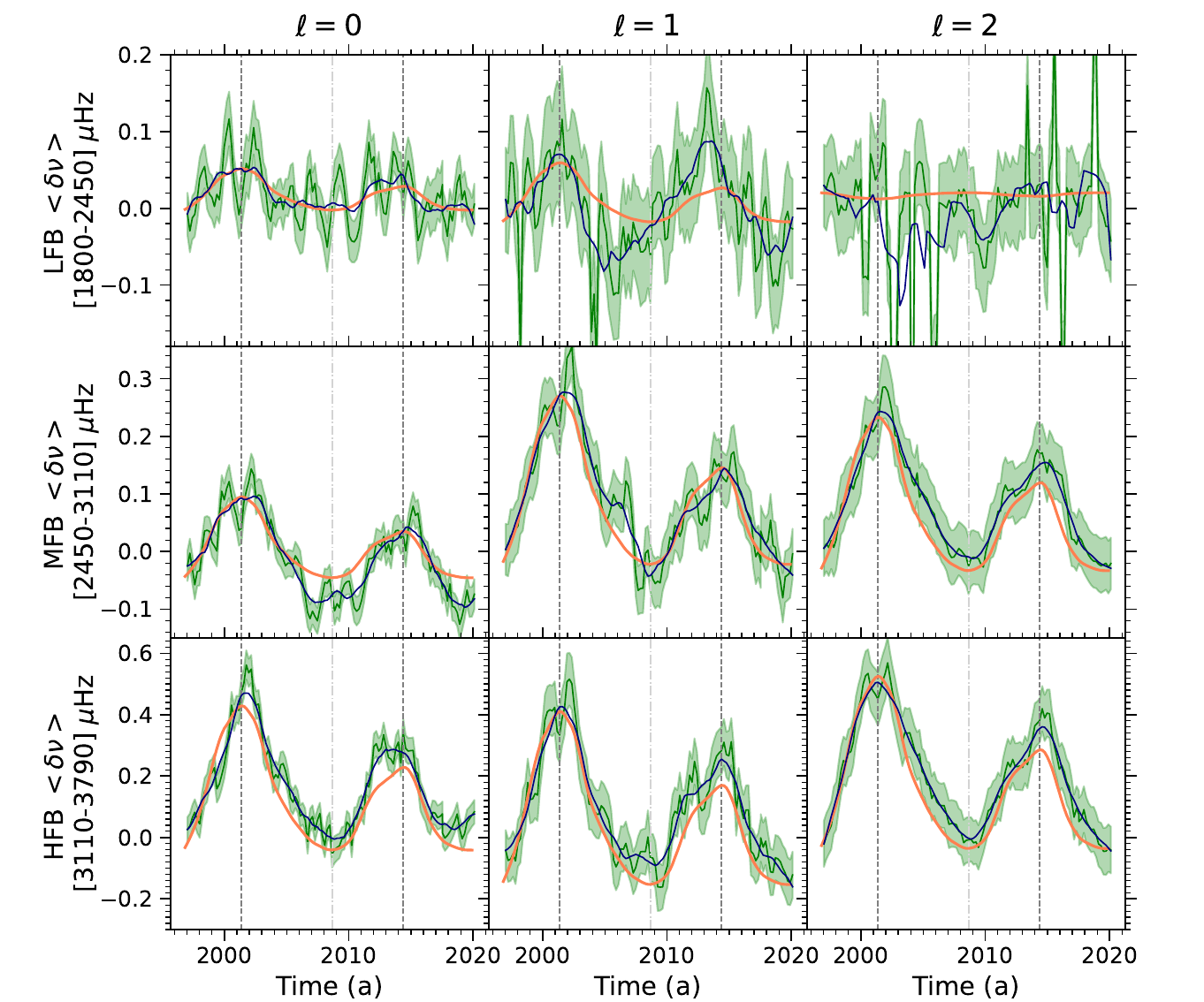}
\end{center}    
    \caption{Temporal evolution of averaged but not smoothed GONG frequency shifts (dark green lines) with errors represented by the light green regions. The signal-to-noise ratio of $\ell$=2 modes in the LFB is too low and the results are not reliable. Same legend as in Fig.~\ref{fig:fig_with_QBO}.}
    \label{fig:fig_GONG}
\end{figure*}

The main differences between the instruments are summarised as follows (see Figs.~\ref{fig:fig_BiSON} and~\ref{fig:fig_GONG}): 
\begin{itemize}
    \item The GONG LFB of $\langle\delta\nu_{\ell=0}\rangle$ shows a clear QBO pattern that is not seen in the Sun-as-a-star instruments. Moreover, the filtered frequency $\langle\delta\nu_{\ell=0}\rangle$ follows perfectly $F_{10.7}$.
\item The BiSON LFB of $\langle\delta\nu_{\ell=0}\rangle$ has the same behaviour as GOLF with an even faster rising phase of cycle 24 and a maximum earlier than in $F_{10.7}$.
\item The BiSON and GONG LFBs of $\langle\delta\nu_{\ell=1}\rangle$ show similar or slightly higher amplitudes of the maximum of cycle 24 compared to cycle 23. 
\item  The GONG LFB of $\langle\delta\nu_{\ell=2}\rangle$ has a very low signal-to-noise ratio. As a consequence, the scaled $F_{10.7}$ appears nearly flat. 
\item The BiSON and GONG MFBs of $\langle\delta\nu_{\ell=1}\rangle$ do not show any excess in cycle 24 near the maximum.

\end{itemize}

Although we note differences between the instruments, we favour GOLF results as the duty cycle of the subseries are in most cases above 95$\%$ while in GONG these vary between 78\% to 92\% and in BiSON these are often below 65$\%$.


\label{Appendix:BiSON_GONG}

\end{appendix}

\end{document}